\documentclass[journal=nalefd,manuscript=letter,layout=twocolumn]{achemso}
\setkeys{acs}{etalmode=truncate,maxauthors=10,articletitle=true} 
\usepackage{graphicx}% Include figure files
\usepackage{dcolumn}% Align table columns on decimal point
\usepackage{bm}% bold math
\usepackage{gensymb}
\usepackage{siunitx}
\usepackage{amsmath}
\usepackage{amssymb}

\newcommand{\dg}[1]{\ensuremath{#1^\circ}}
\newcommand{\wavenumber}[1]{\ensuremath{\SI{#1}{\per\centi\meter}}}
\newcommand{\mumetr}[1]{\ensuremath{\SI{#1}{\micro\meter}}}
\newcommand{\nmetr}[1]{\ensuremath{\SI{#1}{\nano\meter}}}

\author{Nikolai Christian Passler}
 \email{passler@fhi-berlin.mpg.de}
 \affiliation{Fritz-Haber-Institut der Max-Planck-Gesellschaft, Faradayweg 4-6,14195 Berlin, Germany}
\author{I. Razdolski}
 \affiliation{Fritz-Haber-Institut der Max-Planck-Gesellschaft, Faradayweg 4-6,14195 Berlin, Germany}
\author{D. Scott Katzer}
 \affiliation{US Naval Research Laboratory, 4555 Overlook Avenue SW, Washington DC 20375, USA}
\author{D. F. Storm}
 \affiliation{US Naval Research Laboratory, 4555 Overlook Avenue SW, Washington DC 20375, USA}
\author{Joshua D. Caldwell}
 \affiliation{US Naval Research Laboratory, 4555 Overlook Avenue SW, Washington DC 20375, USA}
 \alsoaffiliation{Vanderbilt University, Institute of Nanoscale Science and Engineering, 2201 West End Ave, PMB 350106, Nashville, TN 37235-0106, USA} 
\author{Martin Wolf}
 \affiliation{Fritz-Haber-Institut der Max-Planck-Gesellschaft, Faradayweg 4-6,14195 Berlin, Germany}
\author{Alexander Paarmann}
 \email{alexander.paarmann@fhi-berlin.mpg.de}
 \affiliation{Fritz-Haber-Institut der Max-Planck-Gesellschaft, Faradayweg 4-6,14195 Berlin, Germany}

\title{Second Harmonic Generation from Phononic Epsilon-Near-Zero Berreman Modes in Ultrathin Polar Crystal Films}


\begin{document}

%\begin{tocentry}

%\includegraphics{./TOC_figure.png}

%Some journals require a graphical entry for the Table of Contents.
%This should be laid out ``print ready'' so that the sizing of the
%Inside the \texttt{tocentry} environment, the font used is Helvetica
%8\,pt, as required by \emph{Journal of the American Chemical
%The surrounding frame is 9\,cm by 3.5\,cm, which is the maximum
%permitted for  \emph{Journal of the American Chemical Society}
%graphical table of content entries. The box will not resize if the
%content is too big: instead it will overflow the edge of the box.
%
%This box and the associated title will always be printed on a
%separate page at the end of the document.

%\end{tocentry}

\begin{abstract}
Immense optical field enhancement was predicted to occur for the Berreman mode in ultrathin films at frequencies in the vicinity of epsilon near zero (ENZ). Here, we report the first experimental proof of this prediction in the mid-infrared by probing the resonantly enhanced second harmonic generation (SHG) at the longitudinal optic phonon frequency from a deeply subwavelength-thin aluminum nitride (AlN) film. Employing a transfer matrix formalism, we show that the field enhancement is completely localized inside the AlN layer, revealing that the observed SHG signal of the Berreman mode is solely generated in the AlN film. Our results demonstrate that ENZ Berreman modes in intrinsically low-loss polar dielectric crystals constitute a promising platform for nonlinear nanophotonic applications.

%We report the first experimental observation of resonantly enhanced second harmonic generation (SHG) from a deeply subwavelength-thin aluminum nitride (AlN) film in the mid-infrared. The origin of the SHG yield is the enormous out-of-plane field enhancement of the Berreman mode arising at the longitudinal optic phonon frequency in the vicinity of epsilon near zero (ENZ). Employing a transfer matrix formalism, we show that the field enhancement is completely localized inside the AlN layer, revealing that the observed SHG signal of the Berreman mode is solely generated in the AlN film. Our results demonstrate that ENZ Berreman modes in intrinsically low-loss polar dielectric crystals constitute a promising platform for nonlinear nanophotonic applications. 
\end{abstract}


%%%%%%%%%%%%%%%%%%%%%%%%%%%%%%%%%%%%%%%%%%%%%%%%%%%%%%%%%%%%%%%%%%%%%%%%%%%%%%%%%%
%%%%%%%%%%%%%%%%%%%%%%%%%%%%%%% INTRODUCTION %%%%%%%%%%%%%%%%%%%%%%%%%%%%%%%%%%%%%
%%%%%%%%%%%%%%%%%%%%%%%%%%%%%%%%%%%%%%%%%%%%%%%%%%%%%%%%%%%%%%%%%%%%%%%%%%%%%%%%%

In nanophotonics, nonlinear optical phenomena are driven by the enhancement of local optical fields, which arises due to polaritonic resonances. These are traditionally observed as plasmon polaritons in metallic nanostructures or rough metal surfaces. Such strongly enhanced fields enable a variety of nanoscale applications\cite{Kauranen2012} such as all-optical switching\cite{Lu2011,Ren2011}, low-loss frequency conversion\cite{Sederberg2015,Shibanuma2017}, and highly efficient sensing\cite{Nie1997,Kneipp1997}. In the infrared (IR), an alternative to plasmonic resonances in metals are phonon polaritons supported in polar crystals\cite{Caldwell2015}, as has been demonstrated in various seminal studies\cite{Hillenbrand2002,Wang2013,Caldwell2013,Autore2018}. These phonon polaritons feature longer lifetimes than plasmon polaritons, leading to much larger quality factors and stronger field enhancements\cite{Chen2014,Caldwell2014,Giles2017}, and thus potentially enhanced efficiency of nonlinear optical effects.

One area in nanophotonics of distinct recent interest are investigations of polaritonic modes in plasmonic or polar dielectric subwavelength-thin films that emerge near zero permittivity. Over the past decades, the existence of such thin-film polaritonic modes has attracted broad attention\cite{Ferrell1958,Berreman1963,McAlister1963,Boesenberg1967,Burke1986,Bichri1993,Vassant2012,Newman2015,Nordin2017,Campione2015,Shaykhutdinov2017}. Initially, radiation in a narrow spectral window at the plasma frequency of thin metal films was predicted\cite{Ferrell1958} and later observed\cite{McAlister1963,Boesenberg1967,Burke1986}. Its origin was associated with a collective surface plasma mode with polarization normal to the surface plane of the film\cite{McAlister1963}. At the same time, a similar effect was reported by Berreman in a thin polar dielectric film\cite{Berreman1963}, where absorption occurs at the longitudinal optic (LO) phonon frequency. 

These absorption features in thin films were argued to originate in radiative virtual polaritonic modes\cite{Bichri1993,Vassant2012,Nordin2017}, naturally occurring at frequencies where the real part of the dielectric function crosses zero. This condition is met at the plasma frequency in a metal, and at the LO frequency of a polar crystal film. While these lossy polariton modes disperse on the low momentum side of the light line, it was discovered that a complementary evanescent polariton mode close to the LO frequency of a polar dielectric is also supported outside the light cone\cite{Vassant2012}. Just like the radiative modes, the evanescent polaritons naturally emerge in thin films at frequencies of vanishing dielectric function. Therefore, these modes offer an intriguing platform for exploiting the unique characteristics of waves propagating in so called epsilon near zero (ENZ) materials\cite{Li2015, Liberal2017}. 

While most ENZ studies depend on carefully and intricately designed metamaterials\cite{Joannopoulos2008,Burgos2010,Sakoda2005,Argyropoulos2014}, thin metal or polar dielectric films stand out for their structural simplicity. Previous studies have reported promising applications employing these ENZ polariton modes, such as optoelectronic devices for the ultrafast control of absorption and emissivity\cite{Vassant2012a, Vassant2013,Vasudev2013}, directionally perfect absorption\cite{Luk2014,Feng2012}, or long-range plasmon polaritons for the development of nanophotonic integrated technologies\cite{Berini2001,Berini2009}. 

However, these previous studies mostly focused on the linear optical response, whereas only few reports of the nonlinear conversion efficiency of ultrathin films exist. This efficiency has been proposed to be strongly enhanced at ENZ frequencies\cite{Vincenti2011}, but experimental verification is limited to a few studies of indium tin oxide (ITO) thin layers\cite{Capretti2015,Luk2015,Capretti2015a} excited at frequencies in the near infrared spectral range. The nonlinear optical response of thin films of other materials with phonon resonances in the mid- to far-IR, in particular III-V or III-nitride polar semiconductor compounds, however, has to the best of our knowledge not yet been studied. 

In this work, we investigate the linear and nonlinear optical response of ultrathin ($~\lambda / 1000$) AlN films on a SiC substrate in the radiative regime, where $\lambda$ represents the free-space wavelength at the ENZ condition. We report strong SHG at the AlN LO phonon frequency arising from the Berreman mode in the ultrathin AlN film. The observed SHG yield provides experimental proof of the immense field enhancement inside the film and is attributed to the excitation of the Berreman resonance. Furthermore, we delineate several perspectives based on ENZ polaritons for the deployment of low-loss nonlinear nanophotonic applications.

\begin{figure}
\includegraphics[width=.9\linewidth]{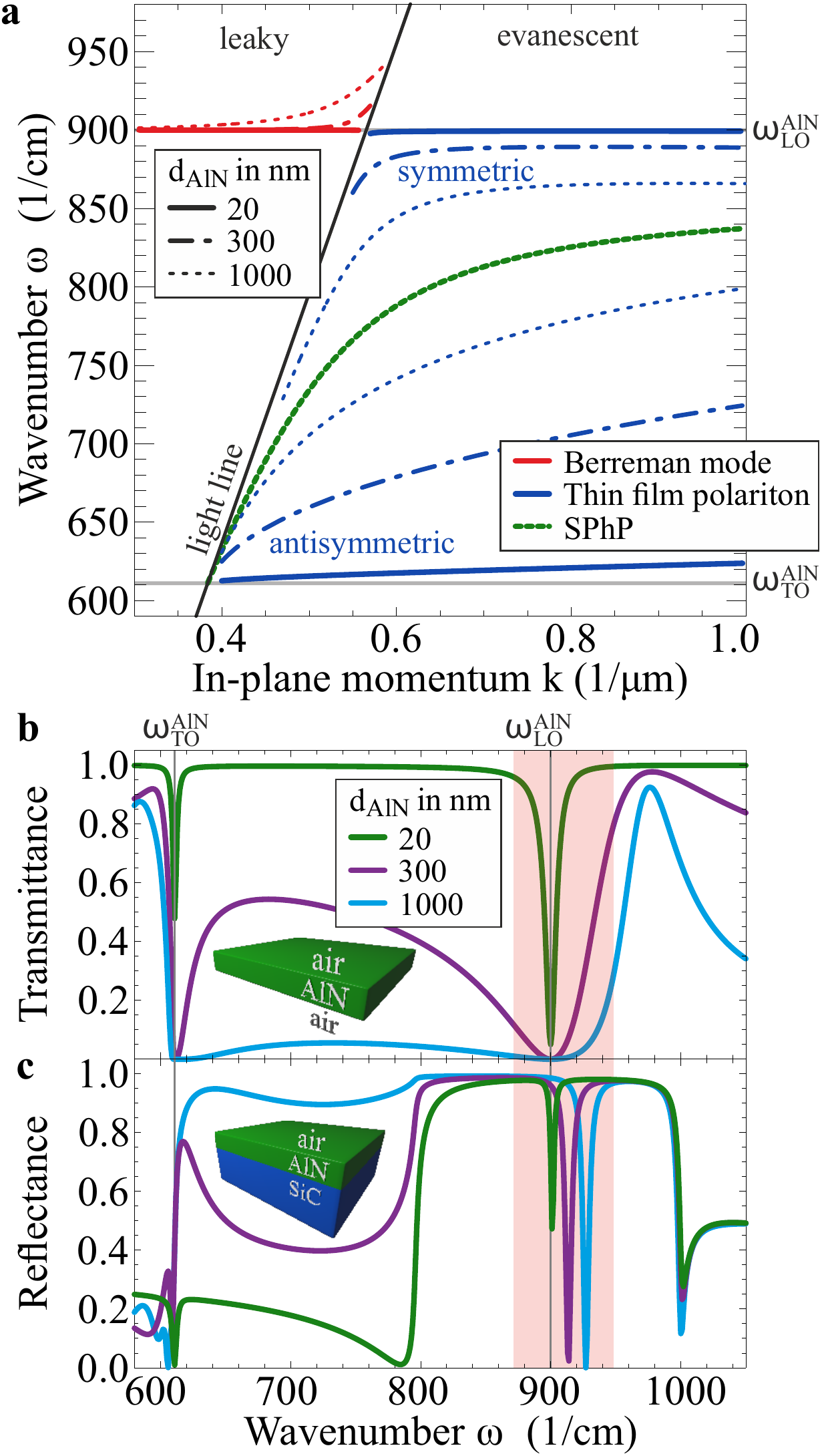}
  \caption{\footnotesize{\textbf{Berreman mode in a freestanding AlN film and on a SiC substrate.} \textbf{a} Calculated dispersions of the Berreman mode and thin film polaritons in a freestanding AlN film, and that of a SPhP at the surface of a bulk AlN crystal. The Berreman mode is flat (red line) for ultrathin films ($d < \lambda/500$), while thicker films (dash dotted/dotted red lines) result in a dispersion bending upwards in the vicinity of the light line in vacuum (black line). Analogous to the symmetric thin film polariton (blue line, upper branch), the dispersion of the Berreman mode lies close to the LO frequency where the real part of the dielectric function exhibits a zero-crossing. \textbf{b} Calculated transmittance of a freestanding AlN thin film for three different film thicknesses $d_{AlN}$ at an incidence angle of \dg{85}. The dip at the LO frequency for the thinnest film corresponds to the Berreman mode (red shade), which disappears with the buildup of the AlN reststrahlen band for increasing film thicknesses. \textbf{c} Calculated reflectance of an AlN thin film on a SiC substrate at an incidence angle of \dg{85}. Here, the Berreman mode appears as a deep dip inside the reststrahlen band of SiC.}}
  \label{fig1}
\end{figure}

%%%%%%%%%%%%%%%%%%%%%%%%%%%%%%%%%%%%%%%%%%%%%%%%%%%%%%%%%%%%%%%%%%%%%%%%%%%%%%%%%%
%%%%%%%%%%%%%%%%%%%%%%%%%%%%%%% THEORETICAL BASICS %%%%%%%%%%%%%%%%%%%%%%%%%%%%%%%%%%%%%%%%
%%%%%%%%%%%%%%%%%%%%%%%%%%%%%%%%%%%%%%%%%%%%%%%%%%%%%%%%%%%%%%%%%%%%%%%%%%%%%%%%%%

A mode in a medium is defined as a solution of Maxwell's equations in the absence of an external perturbation. In a three-layer system, the dispersion of a polaritonic mode can be calculated by numerical evaluation of the following formula:\cite{Raether1988,Burke1986,Campione2015}
\begin{align}
1+\frac{\varepsilon_1 k_{z3}}{\varepsilon_3 k_{z1}}=i \tan{(k_{z2} d)} \left(\frac{\varepsilon_2 k_{z3}}{\epsilon_3 k_{z2}} + \frac{\varepsilon_1 k_{z2}}{\varepsilon_2 k_{z1}}  \right),
\label{eq:disp}
\end{align}

where $\varepsilon$ is the dielectric function, $d$ is the film thickness of layer $2$, $k_{zi}=\sqrt{\frac{\omega^2}{c^2} \varepsilon_{i} - k_{x}^2}$ is the out-of-plane momentum, $k_{x}$ is the in-plane momentum conserved in all layers, and the subscripts $i=1,2,3$ refer to the three layered media. In principle, Eq. \ref{eq:disp} can be solved either for a complex frequency $\omega$ and a real wavevector $k_{x}$, or for a complex $k_{x}$ and a real $\omega$. However, depending on the mode nature and the observables of interest, one of the representations is better suited than the other. We here choose the complex $\omega$ representation following the rationales found in literature \cite{Archambault2009,Campione2015,Vassant2012a,Alexander1974}, especially to account for the virtual nature of the Berreman mode \cite{Kliewer1966a} (for further details, see Supplementary Information Fig. S1).

By solving Eq. \ref{eq:disp} for an air/AlN/air system with varying thickness $d_{AlN}$ of the AlN layer, the dispersion curves shown in Fig. \ref{fig1}a are obtained. On the right hand side of the light line, i.e. in the region of evanescent surface-bound solutions, the symmetric (upper blue lines) and antisymmetric (lower blue lines) thin-film polaritons emerge. Spectrally, these modes are bound inside the AlN reststrahlen region between the TO and LO frequencies $\omega_{TO}^{AlN}$ and $\omega_{LO}^{AlN}$, respectively. For thick films ($d_{AlN}>\mumetr{1}$), the two modes enclose the dispersion of a surface phonon polariton (SPhP) at the interface of a bulk AlN crystal (green line) and, for further increasing film thicknesses, eventually fall back onto this single green curve. On the other hand, for diminishing thicknesses, the two modes are pushed towards $\omega_{TO}^{AlN}$ and $\omega_{LO}^{AlN}$. For $d_{AlN}=\nmetr{20}$ (solid line), the upper mode features a flat dispersion curve in the vicinity of $\omega_{LO}^{AlN}$. At this frequency, the dielectric function approaches zero, and hence the upper mode is an ENZ thin-film polariton.

The Berreman mode (red lines) arises as a continuation on the left hand side of the light line, i.e. in the region of radiative solutions. Contrary to its evanescent counterpart, this leaky polariton mode undergoes a small upward bend close to the light line for larger film thicknesses. For $d_{AlN}=\nmetr{20}$, however, the Berreman mode has a flat dispersion at $\omega_{LO}$ just like the evanescent polariton, and therefore also exhibits ENZ character. 

The feature in the transmittance spectrum of a thin film at the LO frequency originally reported by Berreman\cite{Berreman1963} is reproduced in the transmittance simulations shown in Fig. \ref{fig1}b. The Berreman mode can only be excited by radiation with a non-zero out-of-plane electric field component, and hence $p$-polarized light at oblique incidence is required to observe the absorption peak. A visual explanation for the necessity of $p$-polarized excitation can be found in the spatial electric field distribution of the Berreman mode (see Supplementary Information Fig. 2a and b). Along the in-plane coordinate $x$, the field vectors perform a rotation in the $x$-$z$ plane, i.e. the plane of incidence, with $z$ being the out-of-plane coordinate. In order to accentuate the Berreman absorption feature, the curves shown in Fig. \ref{fig1}b-c were calculated at an incidence angle of \dg{85}, leading to a strongly pronounced dip at $\omega_{LO}^{AlN}=\wavenumber{900}$ for $d_{AlN}=\nmetr{20}$. 

The transition from an ultrathin AlN film with $d_{AlN}=\nmetr{20}$ to a thicker one with $d_{AlN}=\nmetr{1000}$, is characterized by the buildup of the AlN reststrahlen band, featuring vanishing transmittance between the TO and LO frequencies, as shown in Fig. \ref{fig1}b. Because of this buildup, already for $d_{AlN}=\nmetr{300}$, the Berreman absorption dip is strongly broadened and its frequency position is not clearly defined. This is different in Fig. \ref{fig1}c, were we show the reflectance curves for an air/AlN/SiC structure, resembling the experimentally investigated sample. Interestingly, the highly reflective reststrahlen band of the SiC substrate allows to observe the Berreman absorption feature in a reflectance measurement. In contrast to the freestanding film, a sharp and deep minimum is observed even for \nmetr{1000} film thickness, indicating that the Berreman mode is still supported at the thick-film limit. In the Supplementary Information Fig. S3b, the reflectance is shown for film thicknesses up to \mumetr{5}. For $d_{AlN}>\mumetr{1.5}$, the amplitude of the Berreman dip starts to diminish, approaching the optical response of a bulk AlN crystal that does not support a Berreman mode anymore. 

In order to verify that the reflectance dips in Fig. \ref{fig1}c originate in the Berreman mode, the theoretical dispersion for the air/AlN/SiC structure is calculated employing Eq. \ref{eq:disp}. While being quantitatively similar to the air/AlN/air Berreman mode dispersion, the SiC substrate leads to a smaller slope and a reduced frequency in proximity to the light line (see Supplementary Information Fig. S3a). As is shown in Supplementary Information Fig. S3b, the frequencies of the theoretical dispersion at \dg{85} incidence angle and those of the numerical reflectance curves are in excellent agreement, corroborating that a reflectance measurement gives experimental access to the Berreman mode of the air/AlN/SiC structure.

We employ SHG spectroscopy\cite{Paarmann2015} to probe the field enhancement associated with the excitation of the Berreman mode. The strongest field confinement occurs in ultrathin films leading to a strong field enhancement of the Berreman mode, which is a prerequisite for the observation of a significant SHG signal. We therefore focus on two samples with ultrathin AlN films of thickness $d_{AlN}=\nmetr{10}$ and \nmetr{20}. The AlN films were grown by RF-plasma assisted molecular beam epitaxy onto a 4H-SiC substrate, and therefore also feature a hexagonal crystal structure with the $c$-axis being perpendicular to the sample surface. 

%%%%%%%%%%%%%%%%%%%%%%%%%%%%%%%%%%%%%%%%%%%%%%%%%%%%%%%%%%%%%%%%%%%%%%%%%%%%%%%%%%
%%%%%%%%%%%%%%%%%%%%%%%%%%%%%%% METHOD %%%%%%%%%%%%%%%%%%%%%%%%%%%%%%%%%%%%%%%%%%%
%%%%%%%%%%%%%%%%%%%%%%%%%%%%%%%%%%%%%%%%%%%%%%%%%%%%%%%%%%%%%%%%%%%%%%%%%%%%%%%%%%

The reflectance and SHG spectroscopy measurements were performed in a non-collinear autocorrelator setup\cite{Paarmann2015} at \dg{30} and \dg{60} incidence angle employing a tunable, narrow-band, $p$-polarized mid-IR free electron laser (FEL)\cite{Schollkopf2015} as an excitation source. Beforehand, intrinsic higher harmonics of the FEL are blocked by two dichroic \mumetr{7} longpass filters. The reflectance is recorded at \dg{60} by a pyroelectric detector, whereas the two-pulse correlated SHG signal is generated at \dg{45} between the reflected fundamental beams and is measured by a mercury-cadmium-telluride detector. For two $p$-polarized incident beams, the produced SHG signal is also $p$-polarized (PPP configuration). Because of the respective $\chi^{(2)}$ component for $c$-cut crystals being zero, there is no SPP contribution\cite{Paarmann2015,Paarmann2016}. We note that the non-collinear excitation scheme is only applicable for ultrathin films where the shift of the Berreman resonance frequency with incidence angle is negligible (see Fig. \ref{fig1}a), whereas for thicker films, a collinear setup would be necessary\cite{Passler2017}.

%%%%%%%%%%%%%%%%%%%%%%%%%%%%%%%%%%%%%%%%%%%%%%%%%%%%%%%%%%%%%%%%%%%%%%%%%%%%%%%%%%
%%%%%%%%%%%%%%%%%%%%%%%%%%%%%%% EXPERIMENT(S) %%%%%%%%%%%%%%%%%%%%%%%%%%%%%%%%%%%%
%%%%%%%%%%%%%%%%%%%%%%%%%%%%%%%%%%%%%%%%%%%%%%%%%%%%%%%%%%%%%%%%%%%%%%%%%%%%%%%%%%

\begin{figure*}[t]
\includegraphics[width=\linewidth]{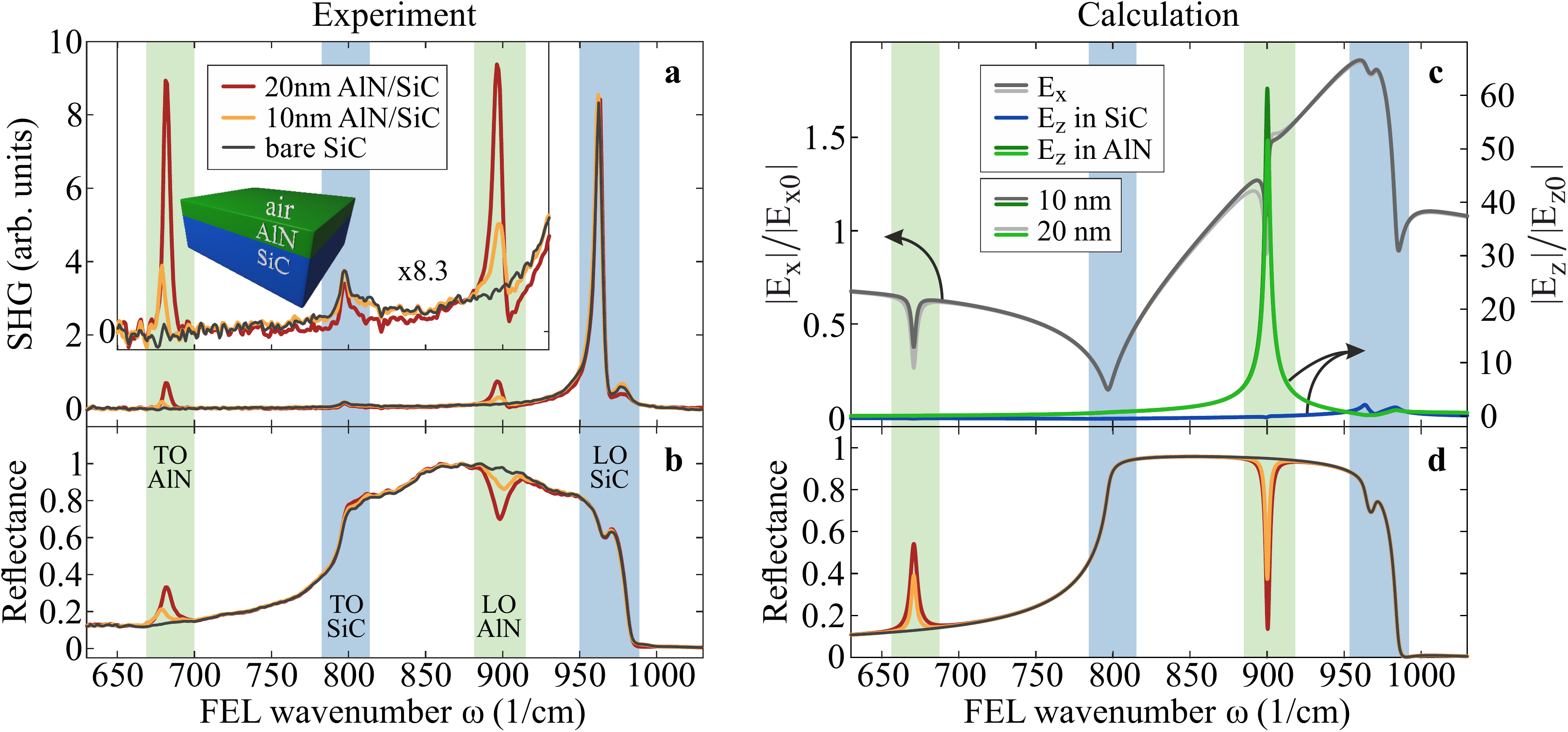}
  \caption{\footnotesize{\textbf{Strongly enhanced SHG from a Berreman mode in AlN.} \textbf{a} and \textbf{b} show the experimental SHG and reflectance spectra, respectively, taken for three samples consisting of (i) a \nmetr{20}, (ii) a \nmetr{10} thin AlN film on a 4H-SiC substrate, and (iii) a bare 4H-SiC crystal. Compared to the reference sample (iii), the AlN thin films only differ at the TO and LO frequencies of AlN, exhibiting small features in the reflectance and a strong SHG signal (enlarged by a factor of 8.3 in the inset in a). The origin of the strong SHG yield is illustrated in \textbf{c}. While the in-plane $E_x$ field enhancement is small at the AlN/SiC interface (grey lines, left y-axis), the out-of-plane $E_z$ fields feature a strong enhancement of $>60$ at the AlN LO frequency (green lines, right y-axis). On the SiC side of the AlN/SiC interface, the $E_z$ fields are small (blue line), revealing that the field enhancement is strictly confined inside the nanometric AlN layer. \textbf{d} shows calculated reflectance curves of the investigated samples, being in excellent agreement with the experiment.}}
  \label{fig2}
\end{figure*}

The experimental SHG and reflectance spectra are plotted in Figs. \ref{fig2}a and b, respectively. There, the yellow and red lines indicate the data for the $10$ and \nmetr{20} thick AlN films, respectively. Additionally, we show spectra for a bare SiC sample (black lines). As has been demonstrated previously\cite{Paarmann2015,Paarmann2016}, the bulk SiC substrate produces SHG peaks at its TO and LO frequencies (blue shades), leading to the same response in all three samples in these regions ($\omega \sim \wavenumber{800}$ and $\omega \sim \wavenumber{970}$). In fact, the only deviations from the bulk reflectance and SHG spectra are seen at the AlN TO and LO frequencies (green shades), where we observe clear, strong peaks in the SHG signal scaling with the AlN film thickness (see inset in Fig. \ref{fig2}a with enlarged vertical axis). The observation of such a sizable SHG yield at the LO frequency is astonishing, especially considering the exceptionally small effective volume of only a few nanometer AlN that is generating the signal.

The SHG intensity $I_{SHG}$ is proportional to the tensor product of the field enhancement $\vec{E}(\omega)$ and the second-order susceptibility tensor $\chi^{(2)}$\cite{Shen1989}:
\begin{align}
I_{SHG}\propto \left| \chi^{(2)}(-2\omega;\omega, \omega) \vec{E}(\omega)\vec{E}(\omega) \right|^2.
\end{align}

It is clear that either a resonance peak in $\chi^{(2)}$ or $\vec{E}(\omega)$ will lead to an enhanced SHG yield and thus a peak in the SHG spectrum. However, at the LO frequency, the second-order susceptibility $\chi^{(2)}$ has no resonances\cite{Paarmann2015}. As for the case of polaritons\cite{Passler2017}, also here we do not need to introduce an additional resonance in the $\chi^{(2)}$ to reproduce the data. We therefore argue that the origin of the reported large SHG yield is the immense electric field enhancement in the AlN thin film. In order to get further insights into the electric field distributions, we employ a $4\times 4$ transfer matrix formalism specifically designed to simultaneously handle media with fully anisotropic as well as isotropic dielectric tensors\cite{Passler2017a}. This allows us to account for the uniaxial anisotropy of both 4H-SiC and hexagonal AlN, leading to an accurate reproduction of the reflectance data with highly detailed qualitative accordance. For instance, even small features like the dip at the high-frequency reststrahlen edge of SiC originating from the SiC anisotropy are accurately reproduced, see Fig. \ref{fig2}d. (Note that for Fig. \ref{fig1} the materials were taken to be isotropic, which is sufficient for the qualitative understanding of the Berreman mode.) 

Quantitatively, the calculations feature a deeper and sharper Berreman dip in the reflectance than in the experiments, see Fig. \ref{fig2}b. This discrepancy is mainly due to growth defects in the AlN layer and an unavoidable strain due to the lattice mismatch between SiC and AlN (1\%)\cite{Tairov1983,Taylor1960}, leading to an effectively increased damping constant of AlN than assumed in the calculations ($\gamma_{AlN}=\wavenumber{2.2}$)\cite{Moore2005}. Furthermore, an additional experimental broadening arises from the FEL linewidth ($\sim \wavenumber{4}$).

In Fig. \ref{fig2}c, we show the in-plane ($E_x$) and out-of-plane ($E_z$) local electric field enhancements at the AlN/SiC interface in both media. Note that $E_x$ and $E_z$ are normalized to their respective incoming field amplitudes $E_{x0}$ and $E_{z0}$. While $E_x$ is conserved at the interface and is generally small (with a maximum value of $\sim1.9$), the $E_z$ field enhancement features a strong peak at $\omega_{LO}^{AlN}$ with a maximum of $>60$ for the \nmetr{10} film. 

Note that while the Berreman reflectance minimum deepens for larger film thicknesses, the $E_z$ field enhancement is already 16\% smaller in the \nmetr{20} than in the \nmetr{10} film. Thus, counter-intuitively, thicker films that feature higher optical absorption, exhibit a smaller degree of field enhancement. In the observed SHG signal, this reduction is compensated by an increasing effective volume, leading to larger SHG yields. However, we emphasize that only ultrathin films ($d_{AlN}<\nmetr{50}$) demonstrate such high field intensities, thus opening new possibilities for deeply subwavelength nanophotonic applications in the IR.

\begin{figure*}
\includegraphics[width=\linewidth]{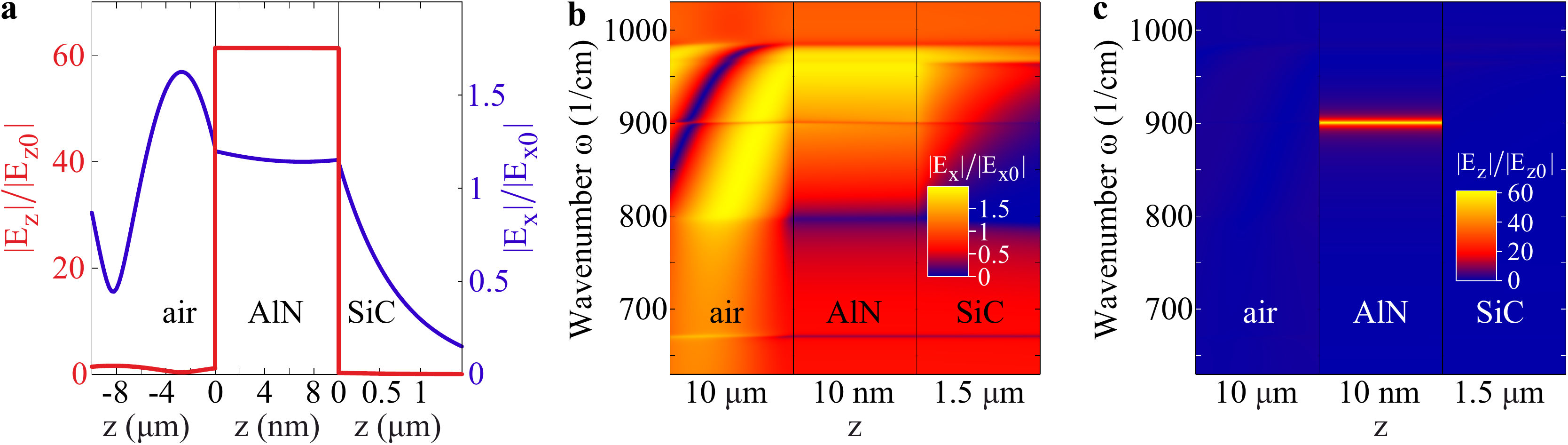}
  \caption{\footnotesize{\textbf{Field enhancement in a 10 nm thin AlN film.} \textbf{a} shows the in-plane and out-of-plane normalized fields $E_x$ and $E_z$, respectively, along the $z$ axis perpendicular to the interfaces and calculated at the Berreman resonance at \wavenumber{900}. The large $E_z$ field enhancement (red) is strongly localized inside the AlN. The $E_x$ field (blue), on the other hand, does not feature any significant field enhancement. Note that in the air layer, the sum of the normalized incoming wave and the reflected wave is plotted, thus resulting in amplitudes larger than 1. \textbf{b, c} Spatio-spectral maps of $E_x$ and $E_z$, respectively. The in-plane field $E_x$ features a small enhancement at the SiC LO (\wavenumber{965}) and a minimum at the SiC TO (\wavenumber{797}). In between in the SiC reststrahlen band, the field decays evanescently into the SiC substrate. Interestingly, a local minimum in $E_x$ can be observed at both the AlN TO (\wavenumber{670}) and the AlN LO (\wavenumber{900}). On the contrary, the out-of-plane field $E_z$ in \textbf{c} features an immense and spectrally sharp field enhancement inside the AlN layer at the AlN LO frequency.}}
  \label{fig3}
\end{figure*}

The $E_z$ field is fully confined inside the AlN layer, which is reflected in the flat frequency dependence and small magnitude of the $E_z$ field enhancement in SiC (blue line in Fig. \ref{fig2}c). This field localization is even better illustrated in Fig. \ref{fig3}a, where we show the spatial distribution of the in-plane $E_x$ and out-of-plane $E_z$ field enhancements as a function of $z$, i.e. along the surface normal, at $\omega=\wavenumber{900}$. The electric field has to obey Maxwell's boundary conditions, i.e. continuity of the in-plane fields ($E_x^{air}=E_x^{AlN}$ and $E_y^{air}=E_y^{AlN}$) and of the out-of-plane displacement field $D_z=\varepsilon E_z$ is required:
\begin{align}
\varepsilon_{air} E_z^{air}=\varepsilon_{AlN} E_z^{AlN}.
\label{contin}
\end{align}
Eq. \ref{contin} is the physical reason for a field enhancement at ENZ conditions, since for vanishing $\varepsilon_{AlN}$ adjacent to air with a finite $\varepsilon_{air}$, the electric field $E_z^{AlN}$ strongly increases in order to fulfill the boundary condition. In a bulk crystal, the field enhancement at ENZ conditions typically reaches values on the order of 1-10 (e.g. in bare SiC\cite{Paarmann2015} or in an AlN/SiC structure, see Supplementary Information Fig. S4a). 

However, in a bulk crystal the phase difference of the incoming and the reflected fields is close to zero, leading to a small total field at the air/AlN interface ($E_z^{air}$) due to destructive interference. As a consequence, following Eq. \ref{contin}, the ENZ induced enhancement of the $E_z^{AlN}$ field is strongly suppressed. In contrast, in the limit of an ultrathin AlN film ($d_{AlN}\lesssim \nmetr{100}$), the phase difference becomes sizable, thus leading to the strong field enhancement as shown in Fig. \ref{fig2}c and Fig. \ref{fig3}a (for details see Supplementary Information Fig. S4b-g).

Fig. \ref{fig3}b and c show spatio-spectral maps of the $E_x$ and $E_z$ fields each normalized to $E_{x0}$ and $E_{z0}$, respectively. Interestingly, $E_x$ features no considerable field enhancement, but exhibits small dips or peaks marking the TO and LO frequencies of both AlN and SiC. The spatio-spectral map of $E_z$ in Fig. \ref{fig3}c, on the other hand, clearly reveals the extreme, spectrally sharp and strongly confined field enhancement in the AlN layer at $\omega=\wavenumber{900}$.

Finally, we turn to the TO frequency of AlN (\wavenumber{670})\cite{Davydov1998}, where the experimental data in Fig. \ref{fig2}a exhibit a strong SHG signal of similar magnitude as at the LO frequency. Quite surprisingly, this AlN TO peak is even larger than the peak at the TO frequency of the SiC substrate. Partially, this can be attributed to a reduced field suppression at $\omega_{TO}$ for thin films compared to a bulk crystal\cite{Paarmann2016} (see Supplementary Information Fig. S5). Notably, the observed peak arises from a resonance in the second-order susceptibility $\chi^{(2)}$ at $\omega_{TO}^{AlN}$, and not from a field enhancement as for the LO peak. Therefore, to fully understand the SHG peak amplitudes at the TO frequencies, a quantitative model of the $\chi^{(2)}$ for AlN would be necessary.

%%%%%%%%%%%%%%%%%%%%%%%%%%%%%%%%%%%%%%%%%%%%%%%%%%%%%%%%%%%%%%%%%%%%%%%%%%%%%%%%%%
%%%%%%%%%%%%%%%%%%%%%%%%%%%%%%% DISCUSSION %%%%%%%%%%%%%%%%%%%%%%%%%%%%%%%%%%%%%%%
%%%%%%%%%%%%%%%%%%%%%%%%%%%%%%%%%%%%%%%%%%%%%%%%%%%%%%%%%%%%%%%%%%%%%%%%%%%%%%%%%%

In this work, we have observed an immense SHG signal arising from a Berreman mode in an ultrathin AlN film excited at ENZ frequencies in the mid-IR. Analogous to previous studies of ITO\cite{Alam2016}, aluminum-doped ZnO\cite{Kinsey2015}, and CdO\cite{Yang2017}, the high optical nonlinearity at ENZ conditions in our system holds high promises for all-optical ultrafast control of polarization switching\cite{Yang2017}, and even over the material's optical properties\cite{Kinsey2015,Alam2016}. However, while all mentioned studies employ ultrathin films excited via free-space radiation -- for which in our system the Berreman mode is accessible, dispersing inside the light cone in vacuum -- a complementary polaritonic ENZ mode exists on the other side of the light line. The linear response of these ENZ polaritons has been studied recently\cite{Passler2018,Runnerstrom2018}, but investigations of their nonlinear response are to the best of our knowledge still lacking. Analogous to the Berreman mode, a strong field enhancement also characterizes the ENZ thin film polariton due to its ENZ environment. We therefore highlight the nonlinear response of ENZ polaritons to be an intriguing subject, specifically in light of the development of polariton-based nonlinear nanophotonics.

Polar crystals such as AlN or SiC, where ENZ conditions are met at the LO phonon resonances in the mid-IR, feature several appealing properties that are unavailable in metals or ITO: (i) The imaginary part of the dielectric function $\varepsilon_2$ at $\omega_{LO}$ ($\varepsilon_{AlN}(\omega=\omega_{LO})=0+0.02 i$) is significantly smaller than in metals\cite{Lynch2012}, and more than one order of magnitude smaller than for ITO ($\varepsilon_{ITO}(\omega=\omega_{LO})=0+0.5 i$)\cite{Capretti2015a}, which strongly increases the field enhancement inside the thin layer and hence the SHG efficiency. (ii) Many polytypes of SiC as well as AlN exhibit a hexagonal crystal structure, resulting in a uniaxial anisotropy of the dielectric tensor. This anisotropy leads to a hyperbolic frequency region between the extraordinary and ordinary LO frequencies, i.e. in the range of the ENZ polaritons, enabling a whole new range of phenomena yet to be explored. These phenomena include, as has been observed in different systems before, negative refraction\cite{RodriguesDaSilva2010}, negative phase velocity\cite{Yoxall2015}, or subdiffraction imaging and focusing\cite{LiCaldwell2015,Dai2015}. (iii) Compared to highly doped semiconductors, one drawback of polar crystals is the lack of tunability of the ENZ frequency, being fixed to the LO phonon. On the other hand, due to relatively short lifetimes of surface plasmon polaritons in metals or most highly doped semiconductors, plasmon-based nanophotonics exhibits intrinsic drawbacks due to inherently high losses, whereas SPhPs in polar crystals feature much longer polariton lifetimes due to long-lived phonon resonances\cite{Caldwell2015,Khurgin2015}. Hence, the employment of the ENZ polariton at the LO frequency offers an appealing alternative for nanophotonic applications, where low-loss ENZ characteristics combine with ultra-high field enhancements.

%%%%%%%%%%%%%%%%%%%%%%%%%%%%%%%%%%%%%%%%%%%%%%%%%%%%%%%%%%%%%%%%%%%%%%%%%%%%%%%%%%
%%%%%%%%%%%%%%%%%%%%%%%%%%%%%%% CONCLUSION %%%%%%%%%%%%%%%%%%%%%%%%%%%%%%%%%%%%%%%
%%%%%%%%%%%%%%%%%%%%%%%%%%%%%%%%%%%%%%%%%%%%%%%%%%%%%%%%%%%%%%%%%%%%%%%%%%%%%%%%%%

In conclusion, we have reported the first observation of a resonantly enhanced SHG yield from a phononic Berreman mode in a deeply subwavelength thin film, exemplified for AlN on a 4H-SiC substrate. The origin of this large SHG signal is the immense out-of-plane field enhancement arising due to the zero-crossing of the dielectric function at the thin film LO frequency, strongly confined to the ultrathin layer. Thanks to low phonon dampings in polar crystals such as AlN and SiC, nanophotonic systems based on such crystals offer an appealing alternative to plasmonics, featuring high-quality resonances with extreme field enhancements. As a possible pathway, we envision ultrathin-film Berreman modes featuring ENZ nature to provide new opportunities for ultrafast all-optical control by taking advantage of the high optical nonlinearity.

\begin{acknowledgement}
We thank Wieland Sch\"ollkopf and Sandy Gewinner for operating the FEL. D.S.K. and D.F.S. acknowledge funding support from the Office of Naval Research. J.D.C. acknowledges financial support from the Office of Naval Research under grant N00014-18-2107 and from Vanderbilt School of Engineering. We thank Christopher J. Winta for careful reading of the paper.
\end{acknowledgement}

\section{Supporting Information}
Theoretical thin film polariton dispersions for either complex frequency or in-plane momentum (Fig. S1), Electric field distribution of the Berreman mode in a \mumetr{1} thick AlN slab (Fig. S2), Comparsion of theoretically and numerically determined Berreman dispersion (Fig. S3), Out-of-plane field enhancement $E_z$ at the interfaces of the air/AlN/SiC system (Fig. S4), Contributions to the SHG yield at the AlN TO frequency (Fig. S5).

\bibliography{berremanSHG}

\end{document}

\title{Supporting Information: \\ Resonant Field Enhancement of Epsilon Near Zero Berreman Modes in an Ultrathin AlN Film}

\author{Nikolai Christian Passler}
 \email{passler@fhi-berlin.mpg.de}
 \affiliation{Fritz-Haber-Institut der Max-Planck-Gesellschaft, Faradayweg 4-6,14195 Berlin, Germany}
\author{I. Razdolski}
 \affiliation{Fritz-Haber-Institut der Max-Planck-Gesellschaft, Faradayweg 4-6,14195 Berlin, Germany}
\author{D. Scott Katzer}
 \affiliation{US Naval Research Laboratory, 4555 Overlook Avenue SW, Washington DC 20375, USA}
\author{D. F. Storm}
 \affiliation{US Naval Research Laboratory, 4555 Overlook Avenue SW, Washington DC 20375, USA}
\author{Joshua D. Caldwell}
 \affiliation{US Naval Research Laboratory, 4555 Overlook Avenue SW, Washington DC 20375, USA}
 \affiliation{Vanderbilt University, Institute of Nanoscale Science and Engineering, 2201 West End Ave, PMB 350106, Nashville, TN 37235-0106, USA} 
\author{Martin Wolf}
 \affiliation{Fritz-Haber-Institut der Max-Planck-Gesellschaft, Faradayweg 4-6,14195 Berlin, Germany}
\author{Alexander Paarmann}
 \email{alexander.paarmann@fhi-berlin.mpg.de}
 \affiliation{Fritz-Haber-Institut der Max-Planck-Gesellschaft, Faradayweg 4-6,14195 Berlin, Germany}

\date{\today}
\maketitle
\beginsupplement

\begin{figure}[t]
\includegraphics[width=1\linewidth]{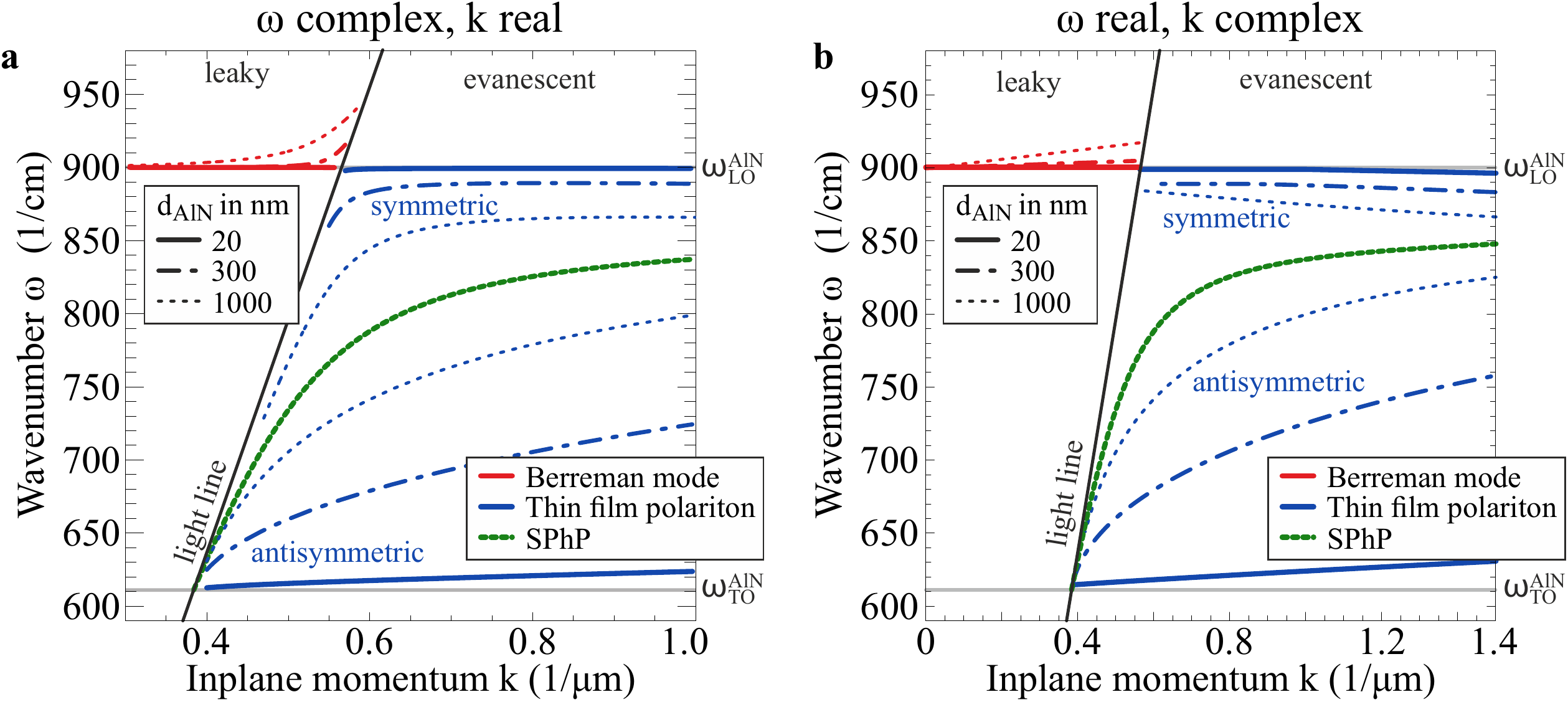}
  \caption{\textbf{Theoretical thin film polariton dispersions for either complex frequency or in-plane momentum.} Solving Eq. 1 of the main text for a complex frequency $\omega$ and real in-plane momentum $k_x$ leads to the dispersions shown in \textbf{a}, while a real $\omega$ and complex $k_x$ yields the curves in \textbf{b}. Depending on the chosen observable, one representation is better suited than the other \cite{Alexander1974}. For instance, the complex frequencies representation is recommended for short pulse excitations, while the complex wavevector representation should be used to analyze the local density of states of a system \cite{Archambault2009}. For the here investigated air/AlN/SiC system, the complex $\omega$ representation (\textbf{a}) is the adequate choice: Firstly, solving the system without damping and thus for real $\omega$ and real $k_x$, qualitatively, the same dispersion curves as shown in \textbf{a} are obtained \cite{Kliewer1966}. Secondly, the leaky Berreman mode resembles a virtual polariton mode, i.e. its fields are related to an energy flow out of the system, which decreases with time \cite{Kliewer1966a}. As has been discussed by Kliewer and Fuchs \cite{Kliewer1966a}, the correct choice for a standard virtual-mode treatment is the complex $\omega$ representation. }
  \label{sfig3}
\end{figure}

\begin{figure}[t]
\includegraphics[width=1\linewidth]{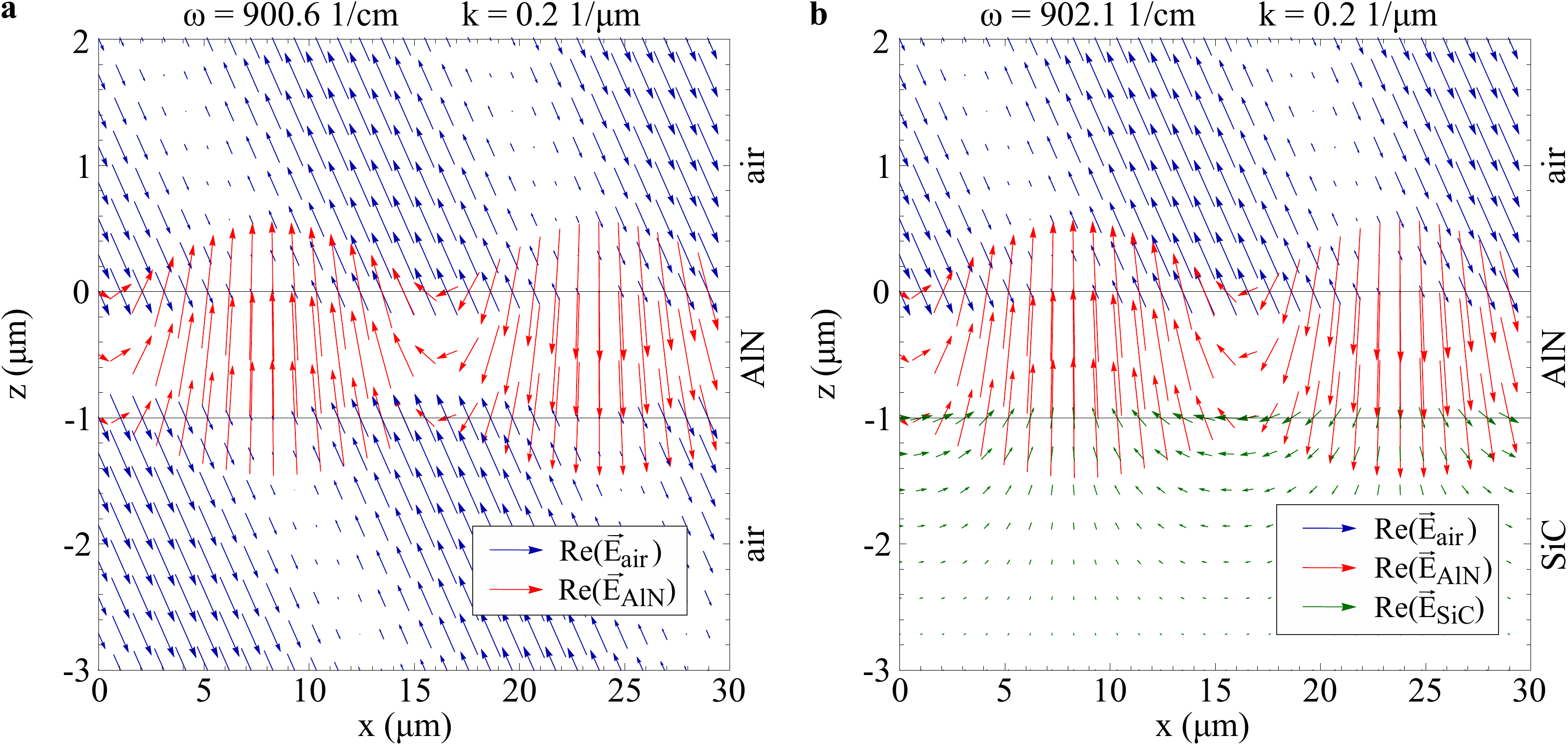}
  \caption{\textbf{Electric field distribution of the Berreman mode in a \mumetr{1} thick AlN slab.} \textbf{a (b)} Vector plots of the electric field distribution in an air/AlN/air (air/AlN/SiC) system of the Berreman mode at an incidence angle of \dg{21} (corresponding to an in-plane momentum of $k=\invmumetr{0.2}$ at $\omega=\wavenumber{900.6}$ ($\omega=\wavenumber{902.1}$), lying on the respective Berreman dispersion). The fields of the Berreman mode (red arrows) perform a counterclockwise rotation in positive x-direction characteristic for polaritonic modes. Please note that the real part is plotted in order to capture the propagation characteristics. The imaginary part features the same behavior with a phase difference of $\pi/2$. The absolute in-plane fields are conserved at the interfaces as required by Maxwells boundary conditions, while the absolute out-of-plane fields experience a strong enhancement in the AlN film. The film thickness was chosen to be \mumetr{1} for illustration purposes. In thinner films, the fields qualitatively exhibit the same distribution, but feature much more pronounced out-of-plane field amplitudes. Interestingly, the substrate material (air in a, SiC in b) has no notable effect on the field distribution of the Berreman mode in the AlN film.}
  \label{sfig2}
\end{figure}

\begin{figure}[t]
\includegraphics[width=1\linewidth]{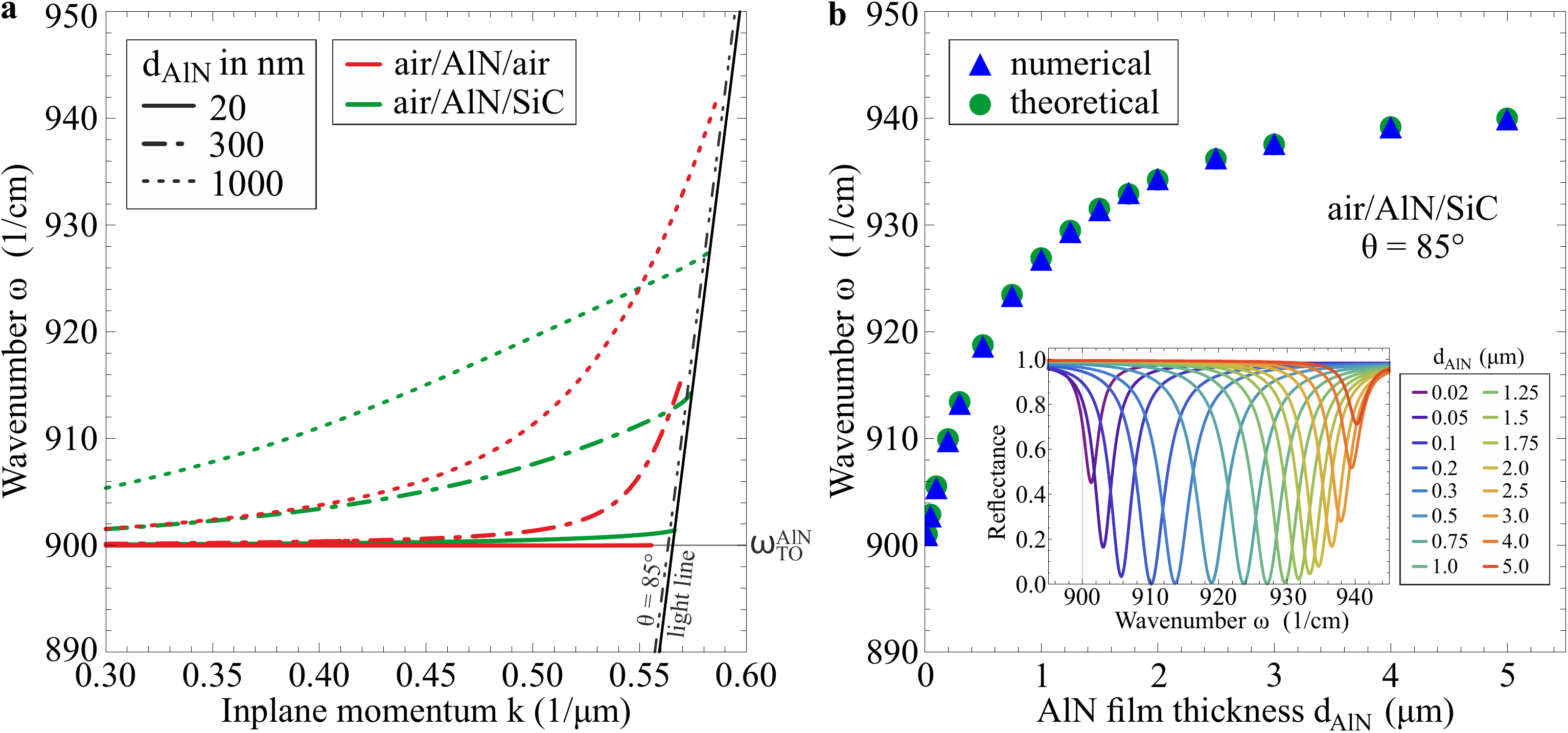}
  \caption{\textbf{Comparison of theoretically and numerically determined Berreman dispersion.} \textbf{a} Berreman dispersions for three different AlN film thicknesses $d_{AlN}$ supported in an air/AlN/air (red curves) and an air/AlN/SiC (green curves) system calculated with Eq. 1 from the main text. The dispersion in a freestanding AlN is flatter for smaller in-plane momenta than for the AlN on a SiC substrate, but exhibits a steeper upwards bending close to the light line. \textbf{b} Plot of the frequencies of the Berreman dispersion as a function of $d_{AlN}$ in the air/AlN/SiC system at an incidence angle of \dg{85}, calculated numerically with the transfer-matrix method (blue triangles) and theoretically as shown in subfigure a (green circles), showing perfect agreement. The inset shows the Berreman reflectance dip as a function of $d_{AlN}$. The frequencies of the minima were extracted to obtain the numerical data (blue triangles).}
  \label{sfig1}
\end{figure}

\begin{figure}[t]
\includegraphics[width=1\linewidth]{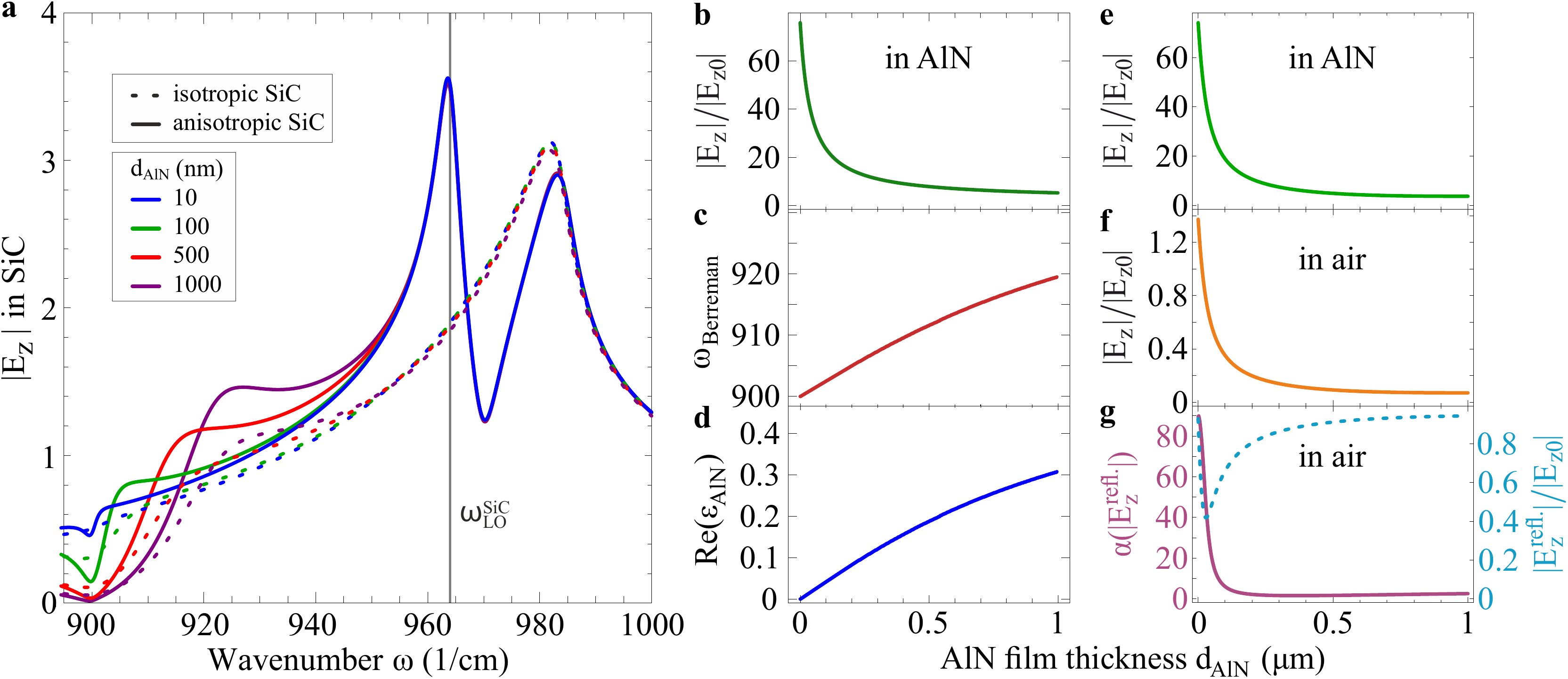}
  \caption{\textbf{Out-of-plane field enhancement  \textit{E\textsubscript{z}} at the interfaces of the air/AlN/SiC system.} \linespread{1}\footnotesize{\textbf{a} $E_z$ field enhancement as a function of frequency in isotropic (dashed lines) and anisotropic (solid lines) SiC for four different AlN film thicknesses $d_{AlN}$ at an incidence angle of \dg{60}. At the LO frequency of anisotropic SiC $\omega_{LO}^{SiC}$, the field enhancement reaches a maximum of $\sim3.5$ due to the zero-crossing of the SiC dielectric function, while the peak at higher frequency arises from a resonance condition in the Fresnel transmission coefficient (also observed in isotropic SiC) \cite{Paarmann2015}. Both features are mainly independent of $d_{AlN}$ because AlN is transparent in this frequency range. The $E_z$ peaks give rise to the SHG signals reported in Fig. 2a of the main text (compare also with \cite{Paarmann2015}). \textbf{b-d} show the $E_z$ field enhancement of the Berreman mode at resonance frequency, the corresponding resonance frequency $\omega_{\text{Berreman}}$, and the real part of the dielectric function of AlN at $\omega_{\text{Berreman}}$, respectively, as a function of AlN film thickness $d_{AlN}$ and at an incidence angle of \dg{60}. Clearly, the $E_z$ field enhancement (b) experiences a strong increase in the limit of ultrathin films ($d_{AlN}<\nmetr{200}$), which can be attributed to the proximity of the Berreman resonance frequency (c) to the ALN LO frequency at \wavenumber{900}, where the dielectric function (d) crosses zero. For thicker films, on the other hand, the Berreman mode disperses at frequencies where $\text{Re}(\varepsilon_{AlN})>0$, yielding a much weaker field enhancement. It is thus seen that only for ultrathin films the Berreman mode, due to its ENZ nature, exhibits the reported strong field enhancement. \textbf{e-g} show the $E_z$ field enhancement at the air/AlN interface in AlN and air, and the reflected wave together with its phase, respectively, as a function of AlN film thickness $d_{AlN}$ at fixed frequency $\omega=\wavenumber{900}$ and at an incidence angle of \dg{60}. Surprisingly, also at fixed ENZ frequency, the field enhancement in AlN (e) decreases with increasing film thickness, which, following Eq. 3 of the main text, happens because of a decreasing $E_z$ field in air (f). The latter behavior arises due to the following reasons. The total field in air is the difference of the incident and the reflected field, $E_z = E_z^{\text{inc.}}-E_z^{\text{refl.}}$, where $E_z^{\text{inc.}}\equiv 1$ and its phase $\alpha(E_z^{\text{inc.}})=0$. The reflected field $E_z^{\text{refl.}}$ has a minimum value of $\sim 0.4$ at $d_{AlN}=\nmetr{30}$ due to the Berreman resonance. Its phase, however, approaches \dg{90} in the limit of a vanishing AlN film, but drops rapidly to values close to \dg{0} for increasing film thickness. As a consequence, for a phase difference of \dg{0}, the reflected field and the incoming field interfere destructively at the air/AlN interface, leading to a small total field amplitude for increasing film thickness both in air and in AlN. In the ultrathin film limit, on the other hand, the significant phase difference of incoming and reflected wave leads to a sizable total field in air, and thus, in combination with the ENZ condition (see Eq. 3 of the main text), to the strong $E_z$ field enhancement inside the AlN film.}}
  \label{sfig3}
\end{figure}

\begin{figure}[t]
\includegraphics[width=1\linewidth]{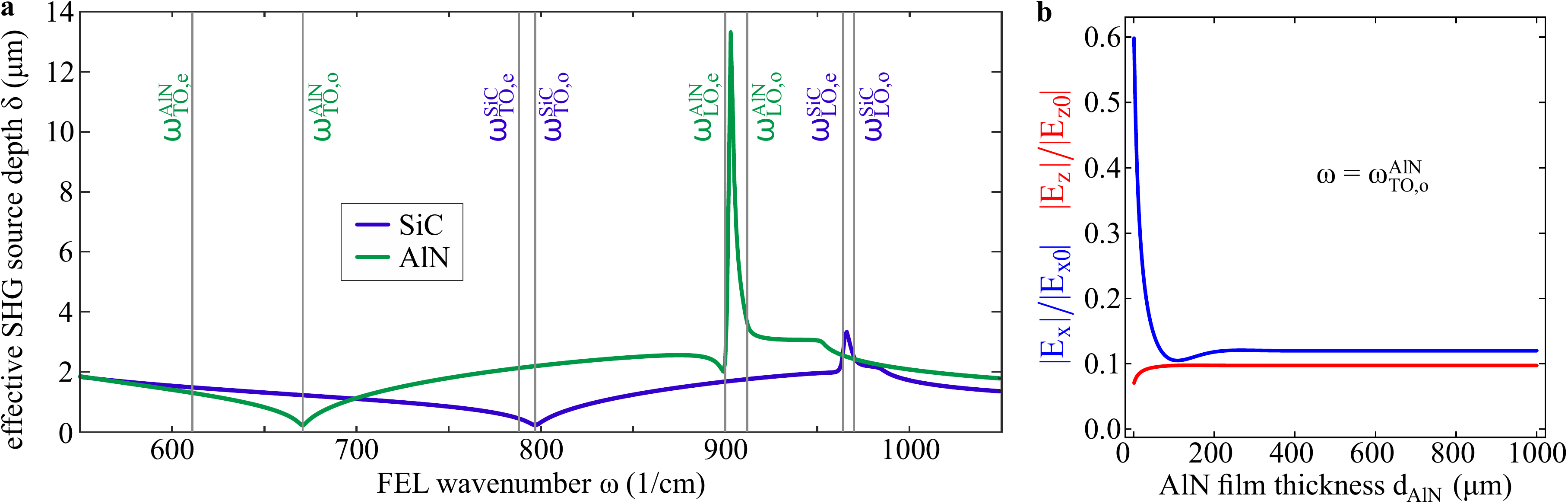}
  \caption{\textbf{Contributions to the SHG yield at the AlN TO frequency.} \textbf{a} Effective SHG source depth $\delta = 1/\Delta k$ for bulk SiC and bulk AlN calculated by means of the wavevector mismatch $\Delta k^2= \left| \vec{k}_{\text{SHG}}-\vec{k}_{\text{1,in}}-\vec{k}_{\text{2,in}} \right|^2$.\cite{Paarmann2016} At the ordinary TO frequency, $\delta$ features a minimum of $\sim\nmetr{200}$ whereas in the range of the LO frequencies, $\delta$ reaches maximum values of a couple of micrometer. Thus, in a bulk sample, a lot more volume contributes to the SHG signal at the LO than at the TO, whereas in a very thin film ($d< \nmetr{200}$), the effective source depth is equal at both frequencies and given by the film thickness. Therefore, in an ultrathin AlN film, the TO SHG signal is stronger compared to the LO signal than for a bulk crystal. \textbf{b} In-plane $E_x$ and out-of-plane $E_z$ fields at the ordinary AlN TO frequency (\wavenumber{670.8}) as a function of AlN film thickness. As stated in the main text, in a bulk crystal the field enhancement is strongly suppressed at the TO frequency\cite{Paarmann2016}, leading to a suppressed SHG yield despite the strong $\chi^{(2)}$ resonance. In an ultrathin film, however, this suppression is reduced, as can be seen by the sharp increase of the $E_x$ field for $d_{AlN}<\nmetr{100}$.}
  \label{sfig3}
\end{figure}

\clearpage

\bibliography{supplMat}